\documentclass{compactarticle}
\usepackage[utf8]{inputenc}
\usepackage{amsmath}
\usepackage{comment}
\usepackage[version=4]{mhchem}
\usepackage{cleveref}
\usepackage{booktabs}
\usepackage{tabulary}
\usepackage{multirow}
\usepackage[normalem]{ulem}

\title{Dissecting flux balances to measure energetic costs in cell biology: techniques and challenges}

\author
{
Easun Arunachalam$^{1,\dagger,\ast}$, William Ireland$^{2, \dagger}$, Xingbo Yang$^{1, \dagger}$, and Dan Needleman$^{1,2,3}$\\
\\
\normalsize{$^{1}$Department of Molecular and Cellular Biology, Harvard University, Cambridge, MA, USA}\\
\normalsize{$^{2}$School of Engineering and Applied Sciences, Harvard University, Cambridge, MA, USA}\\
\normalsize{$^{3}$Center for Computational Biology, Flatiron Institute, New York, NY, USA}\\
\normalsize{$^\dagger$These authors contributed equally to this work}\\
\normalsize{$^\ast$To whom correspondence should be addressed: arunachalam@g.harvard.edu}
}

\begin{document}

\maketitle

\section*{Abstract}
Life is a nonequilibrium phenomenon: metabolism provides a continuous supply of energy that drives nearly all cellular processes. However, very little is known about how much energy different cellular processes use, i.e. their energetic costs.  The most direct experimental measurements of these costs involve modulating the activity of cellular processes and determining the resulting changes in energetic fluxes. In this review, we present a flux balance framework to aid in the design and interpretation of such experiments, and discuss the challenges associated with measuring the relevant metabolic fluxes. We then describe selected techniques that enable measurement of these fluxes. Finally, we review prior experimental and theoretical work that has employed techniques from biochemistry and nonequilibrium physics to determine the energetic costs of cellular processes.

%Table of Contents
\tableofcontents

\section{Introduction}

Cells consume energy derived from the environment to power diverse processes including biosynthesis, ion pumping, cytoskeletal remodeling, signal transduction, error correction and information processing. While the metabolic enzymes and pathways that transduce energy have been extensively studied, very little is known about how much energy is used by different cellular processes. Developing a quantitative understanding of energy usage in cell biology - i.e. cell biological energetic costs - would have broad implications beyond energy metabolism. Since the rate of energy usage is a measure of the combined rates of all energy-consuming reactions, establishing the energetic costs of cellular processes will provide a powerful means to characterize their systems-level behavior \textit{in situ}. Not only will this be useful for testing mechanistic models of these processes, it will also help answer fundamental questions concerning their efficiency and the potential impact of constraints from nonequilibrium thermodynamics \cite{yang2021physical}. Knowing the energetic costs of cellular processes may also provide insight into the changes in metabolism that take place over development or as result of diseases \cite{yang2021physical}, and may have implications for the evolution of these processes \cite{lynch2015bioenergetic}.

Adenosine triphosphate (ATP) is the primary ``energy currency'' in cells: transduction of free energy from the breakdown of nutrients enables the synthesis of ATP from adenosine diphosphate (ADP) and phosphate (P$_\text{i}$), and the hydrolysis of ATP to ADP releases free energy that is used to power many cellular processes. The rate at which ATP molecules are hydrolyzed to supply free energy for a process is often referred to as the energetic cost of that process.

Unfortunately, it is not currently possible to directly measure the rate at which individual cellular processes consume ATP \textit{in situ}. Thus, all of our knowledge of the energetic costs of cell biological processes comes from estimates and indirect measurements. Estimates of energetic costs of particular processes typically combine knowledge of the individual chemical steps in those processes, the ATP stoichiometry of those chemical reactions, and inferences of the total rate of those processes (based on cellular composition, growth rate, and other physiological measurements) \cite{forrest1971generation,stouthamer1973theoretical,lynch2015bioenergetic}. While such estimates have been very informative, they are based on numerous assumptions whose validity is difficulty to verify. There have only been limited efforts to experimentally test such estimates of energetic costs, and when apparent disagreements are found \cite{waterlow1989energy,wieser2001hierarchies}, it can be challenging to determine if this reflects limitations of the estimates, the measurements, or both.

Experimental measurements of energetic costs are often based on modulating the activity of target processes and quantifying concomitant changes in total cellular energy usage \cite{smith1995protein,buttgereit1995hierarchy}. Typically, in such studies a process of interest is inhibited, the change in total cellular metabolic flux is measured, and the inferred change in ATP consumption rates is interpreted as the energetic cost of the inhibited process. One major concern highlighted by these and other studies is the fact that the activity of different cellular processes is inextricably linked, making their respective costs challenging to dissect experimentally \cite{wieser2001hierarchies}. Furthermore, the scope of these studies has been limited; most cellular processes in systems of interest have not yet been studied systematically through such experiments.

Despite their drawbacks, activity modulation experiments remain the most direct method to probe the energetic costs of cellular processes. In this review, we show how theory based on coupled flux balances can be used to interpret such experiments and infer the energetic costs of cellular processes. The balance equations can inform the design of activity modulation experiments by enumerating the individual fluxes that must be quantified for this analysis, and provide a framework to infer energetic costs from changes in the coupled fluxes. Additionally, we use this framework to highlight the challenges associated with interpreting changes in fluxes. We then discuss the strengths and limitations of the different technologies used to measure bioenergetic fluxes and characterize their coupling, and review studies that have employed these approaches to measure energetic costs.

\section{Coupled flux balances are a unifying framework for interpreting bioenergetic measurements}
\label{sec:balances}

The ATP consumption rates of individual cellular processes of interest cannot be directly measured. While challenging to determine, the total combined ATP consumption rate of all processes in a cell, is, in principle, experimentally accessible. This suggests an indirect approach to measuring the ATP consumption rate of a process of interest: change the activity of the processes, determine the resulting change in the total ATP consumption rate, and use these changes to calculate the energetic cost of the process. While this is a promising and powerful approach, there are numerous subtleties associated with using changes in metabolic fluxes to infer energetic costs. In what follows, we will use a systematic categorization of the relationships between the rates of different ATP-producing and ATP-consuming reactions to help interpret such changes in metabolic fluxes.

An ATP mole balance on the system of interest relates changes in ATP levels to fluxes through ATP-producing and ATP-consuming reactions. While the fluxes through many of these ATP-dependent reactions are not directly accessible, they are linked to other reactions in the cell whose rates may be straightforwardly measured. Therefore, careful consideration of the ATP mole balance, as well as of the coupled mole balances on other chemical species and the energy balance, provides a means to calculate and interpret changes in ATP fluxes. These balance equations inform experimental design by indicating which experimentally accessible reaction rates must be measured in order to infer the ATP consumption rate of a given process from an activity modulation experiment. 

We therefore begin by considering the general mole balance for a chemical species $ i $  in the cell. $n_i$, the number of moles of this species in the cell, can change due to the species being imported (with flux $ J_{i,\text{import}} $) or exported (with flux $ J_{i,\text{export}} $), or due to the species being produced or consumed by reaction $ k $ (with flux $ J_{ik}$). Summing over all reactions involving this species gives:

\begin{align}
    \dfrac{dn_i}{dt} &= J_{i,\text{import}} - J_{i,\text{export}} + \sum_{\text{rxns } k} J_{ik}
    \label{eq:fluxes1a}
\end{align}

\noindent Note that $ J_{ik} $ is negative for reactions that are net consumers of species $ i $. We next consider the mole balance for ATP. Import and export of ATP can be safely neglected, so enumerating all reactions that produce or consume ATP gives:

\begin{align}
    \dfrac{dn_\text{ATP}}{dt} &= \sum_{\text{ATP-prod rxns } k'} J_{\text{ATP},k'} + \sum_{\text{ATP-cons rxns } k''} J_{\text{ATP},k''}
    \label{eq:fluxes1b}
\end{align}

\noindent Here, we have divided the reactions into ATP-producing reactions (with positive fluxes) and ATP-consuming reactions (with negative fluxes). Such a division is general because, while the flux of some reactions may change signs under different conditions, at any given time the flux of all reactions is either positive or negative (or zero). In many systems of interest, glycolysis and respiration are the dominant ATP producing pathways, in which case  $\sum_{\text{ATP-prod rxns } k'} J_{\text{ATP},k'} = J_\text{ATP,glycolysis} + J_\text{ATP,respiration}$.

\begin{figure}[!htpb]
    \centering
    \includegraphics[width=\textwidth]{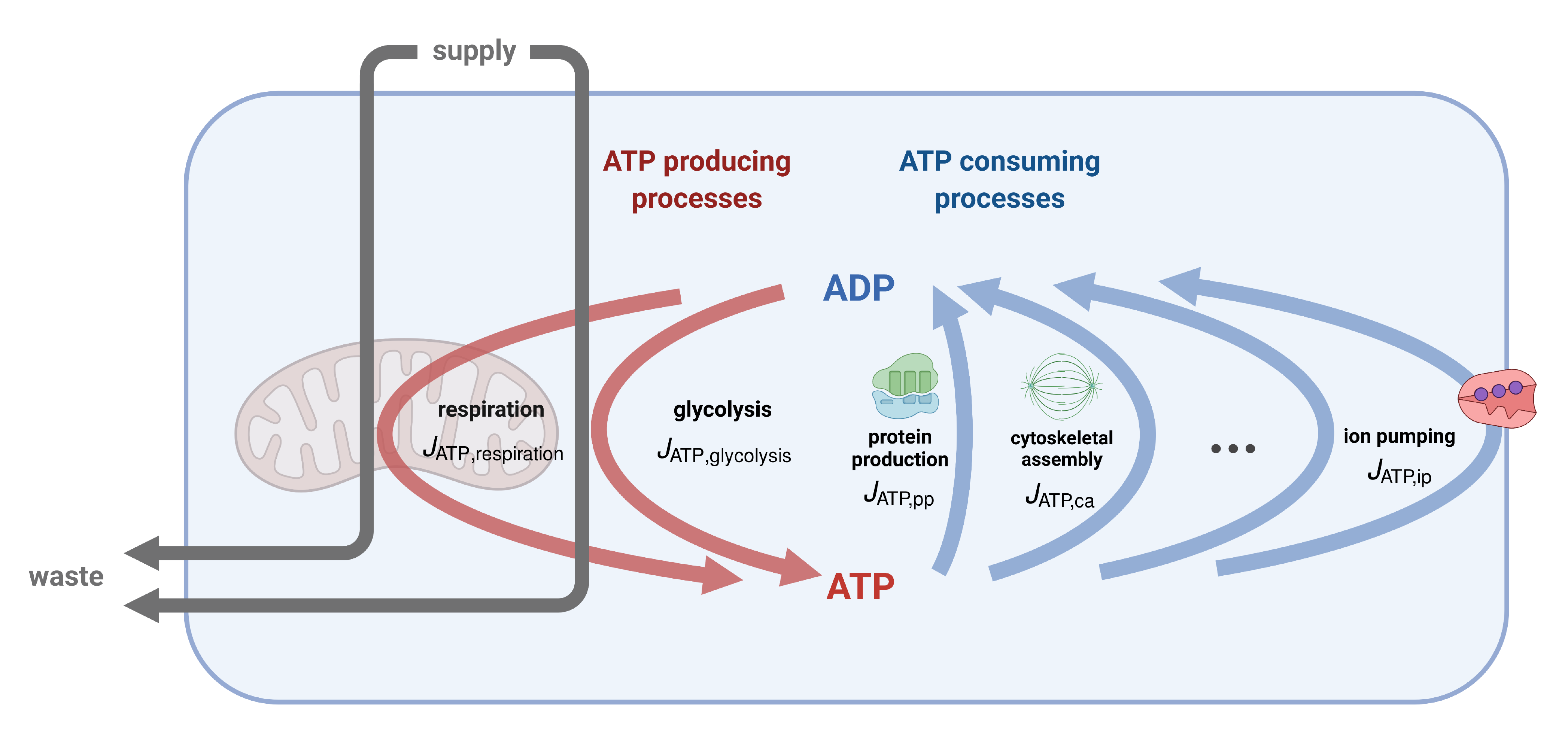}
    \caption{
        \textbf{Many individual processes and pathways contribute to the total ATP production and consumption fluxes, which are balanced at steady state.}
        ATP production, primarily by glycolysis and respiration (red arrows), is balanced by consumption of ATP by numerous cellular processes, including protein production, cytoskeletal assembly, and ion pumping (blue arrows) at steady state. ATP-producing pathways are powered by catabolism of carbon-containing molecules supplied to cells (gray arrows). Portions of this figure were created with BioRender \cite{biorender}.
    }
    \label{fig:overview}
\end{figure}

At steady state, $ \frac{dn_\text{ATP}}{dt} = 0$, and the ATP production and consumption fluxes are balanced. The ATP consuming reactions in this balance can be conceptualized as containing contributions from numerous cell biological processes, each with their own coarse-grained ATP-consuming flux, such as $k''$ = protein production, $k''$ = cytoskeletal assembly, and $k''$ = ion pumping (Fig. \ref{fig:overview}). These processes can be further decomposed into more elementary processes, which in turn can be further decomposed, all the way down to individual chemical reactions that hydrolyze ATP.

The energetic cost of a cell biological process $k''$ is its corresponding ATP consumption flux, $J_{\text{ATP},k''}$. It is not generally possible to directly measure such fluxes. However, at steady state the total ATP consumption flux equals the total ATP production flux, so measuring the total rate of ATP synthesis enables inference of the total rate of ATP hydrolysis. A complicating factor is that direct measurement of ATP synthesis rate is not always possible either. While total cellular ATP synthesis rates have been directly measured by spectroscopic methods \cite{brown197731p}, these techniques are not widely used in modern bioenergetics studies and may be difficult to apply to many systems. Thus, the methods most commonly used to resolve ATP synthesis fluxes rely on coupling between ATP-producing reactions and other reactions in the same pathway, or other processes whose rates can be more easily measured. For example, ATP synthesis fluxes are directly linked to carbon fluxes (as illustrated in Fig. \ref{fig:overview}) and oxygen consumption rate (a measure of oxidative phosphorylation) which are both experimentally accessible. Because the structure of these biochemical pathways is known, measurements of these coupled fluxes enable calculation of $ J_\text{ATP,glycolysis} $ and $ J_\text{ATP,respiration}$, as discussed in Sec. \ref{sec:glycresp}.

The simplest case of these balances is when a cell is at steady state, not changing and not growing (e.g. cell cycle-arrested oocytes). In this case, the time derivatives are eliminated, and ATP production and consumption are balanced. A slightly more complex case entails cells that grow at a constant rate without changing their overall composition (e.g. microbes in balanced growth). It is straightforward to show that in such case normalization by the mass (i.e. considering ATP per unit biomass) again yields a steady-state system with zero time derivatives. The most complicated, general case entails explicit time dependence, e.g. in the case of developing embryos which both grow and change their composition. In such systems, dynamic changes in molar composition, production, and consumption may be non-negligible and must be measured in order to infer ATP synthesis/hydrolysis fluxes. For the moment, we shall focus on steady-state systems, where the total ATP production flux is equal to the total ATP consumption flux.

In steady-state systems, overall changes in ATP consumption rate are compensated by corresponding changes in production rate. Thus, manipulating the rate of a single ATP-consuming process of interest and measuring the concomitant change in ATP production is an intuitively appealing experiment to infer the energetic cost of that process. Let us consider a hypothetical experiment where the rate of a given ATP-consuming process of interest $ p $ was decreased by 50\% while leaving the rates of all other ATP-consuming processes $k' \neq p$ unchanged, and it was observed that the global ATP production rate of the system decreased by some amount $ x $ mol ATP/s. Assuming the energetic cost of a process is linear in its rate, the naive interpretation would be that the energetic cost of process $ p $ was $ 2 x $ mol ATP/s.

In reality, the potential coupling between different cellular processes can complicate the interpretation of such experiments. Processes can be directly coupled, e.g. through signaling pathways that regulate their activity, and indirectly coupled, e.g. through shared pools of substrates or cofactors. Altering the activity of certain processes can cause changes in physiology which affect the rates of a multitude of other cellular activities, and therefore their respective rates of ATP production or consumption. For example, manipulating the activity of a process may alter cells' growth rates, changing the rates of many biosynthetic reactions. 

Additionally, straightforward interpretation of a change in ATP production rates as the energetic cost of the targeted process requires that the total ATP consumption of the cell vary linearly with the ATP demand of that process. It is not yet clear whether this is generally the case, and systematic studies are needed to answer this question in each system of interest. There is reason to believe that this could complicate our understanding of flux changes: the enzyme adenylate kinase, which catalyzes the reversible reaction \ce{ATP$+$AMP <-> 2 ADP}, operates close to equilibrium and is known to buffer the energy charge of the cell (([ATP]+1/2[ADP])/([ATP]+[ADP]+[AMP])) \cite{delafuente2014}. Other ATPases can have a Michaelis-Menten constant $K_M$ close to intracellular ATP/ADP levels, and their reaction rates and directions could be similarly sensitive to changes in ATP/ADP induced by perturbations \cite{buttgereit1995hierarchy,park2016metabolite}. Their contribution to overall ATP production and consumption fluxes could be different in systems before and after process activity changes. This would introduce nonlinearities into the relationship between the ATP demand of a process of interest and the global ATP consumption of the system, hampering interpretation of activity modulation experiments. The extent of this effect is unknown.

Thus, it is important to characterize the dependence of different processes' ATP consumption rates on ATP levels (and possibly levels of ADP, AMP, phosphate, and other species which affect their rates) to properly infer the cost of a single process of interest. One way to conceptualize the impact of this complication is to divide ATP consuming processes into those whose fluxes are sensitive to levels of these species, and those whose fluxes are insensitive to levels of these species. Reactions which respond to ATP levels (or the levels of ADP, AMP, phosphate, etc.) will tend to buffer ATP levels (``buffer'' processes) by two possible mechanisms: 1) some ATPases will increase their rate of consumption as ATP levels rise and decrease as they fall, and 2) others can switch their flux from consumption to production depending on the levels of ATP, ADP, AMP and other species. In contrast, many other processes are relatively insensitive to ATP levels and continue to consume ATP at nearly constant rates (``demand'' processes). This can be formalized by separating the ATP-consuming fluxes in Eq. \ref{eq:fluxes1b} on the basis of their sensitivity to ATP levels over the range explored in a given modulation experiment:
\begin{align}
    \dfrac{dn_\text{ATP}}{dt} &= \underbrace{\left( J_\text{ATP,glycolysis} + J_\text{ATP,respiration} \right)}_{\text{ATP production}} - \underbrace{ \left( \sum_{\text{rxns } k'} J_{\text{ATP demand},k'} + \sum_{\text{rxns } k''} J_{\text{ATP buffer},k''} \right)}_{\text{ATP consumption}}
    \label{eq:fluxes1c}
\end{align}

Here, $ J_{\text{ATP demand},k'} $ is the ATP consumption flux associated with ATP level-insensitive process $ k'$ and $ J_{\text{ATP buffer},k''} $ is the ATP production flux associated with the buffering process $ k'' $; these fluxes are defined to be positive when proceeding in the direction of ATP synthesis. The partitioning of the reactions in the cell into the two sums depends on the $ K_M $ for ATP of the machinery in each process and the range of ATP concentrations explored in a given experiment, which together determine which reaction rate changes can be safely neglected.

In summary, the ATP consumption rates of individual processes cannot be directly measured. However, Eq. \ref{eq:fluxes1c} shows how total ATP production and consumption by all processes is balanced in steady-state systems. This structure motivates the use of activity modulation experiments to probe the energetic costs of individual processes. Despite the complications associated with the interpretation of ATP flux changes, these experiments are the most direct method to determine energetic costs. Therefore, in the following section we discuss how the relationship between Eq. \ref{eq:fluxes1c} and other mole balances enables determination of ATP production rates, and therefore the energetic costs of different processes, from activity modulation experiments. We then consider the caveats of these measurements, and subtleties of activity modulation experiments, including coupling between different fluxes.

\section{Chemical and calorimetric methods to calculate ATP production fluxes}

Inferring the energetic cost of a specific cellular process requires measuring changes in ATP production fluxes caused by modulating the activity of that process. The major ATP-producing pathways in many systems are glycolysis and respiration, whose rates can be measured by several different experimental methods. In this section, we will discuss the logic underlying chemical and calorimetric methods to measure these fluxes, and the respective challenges of these methods.

\subsection{Inferring ATP production rates by chemical measurements of glycolytic and respiratory fluxes}
\label{sec:glycresp}

\begin{figure}[!htpb]
    \centering
    \includegraphics[width=\textwidth]{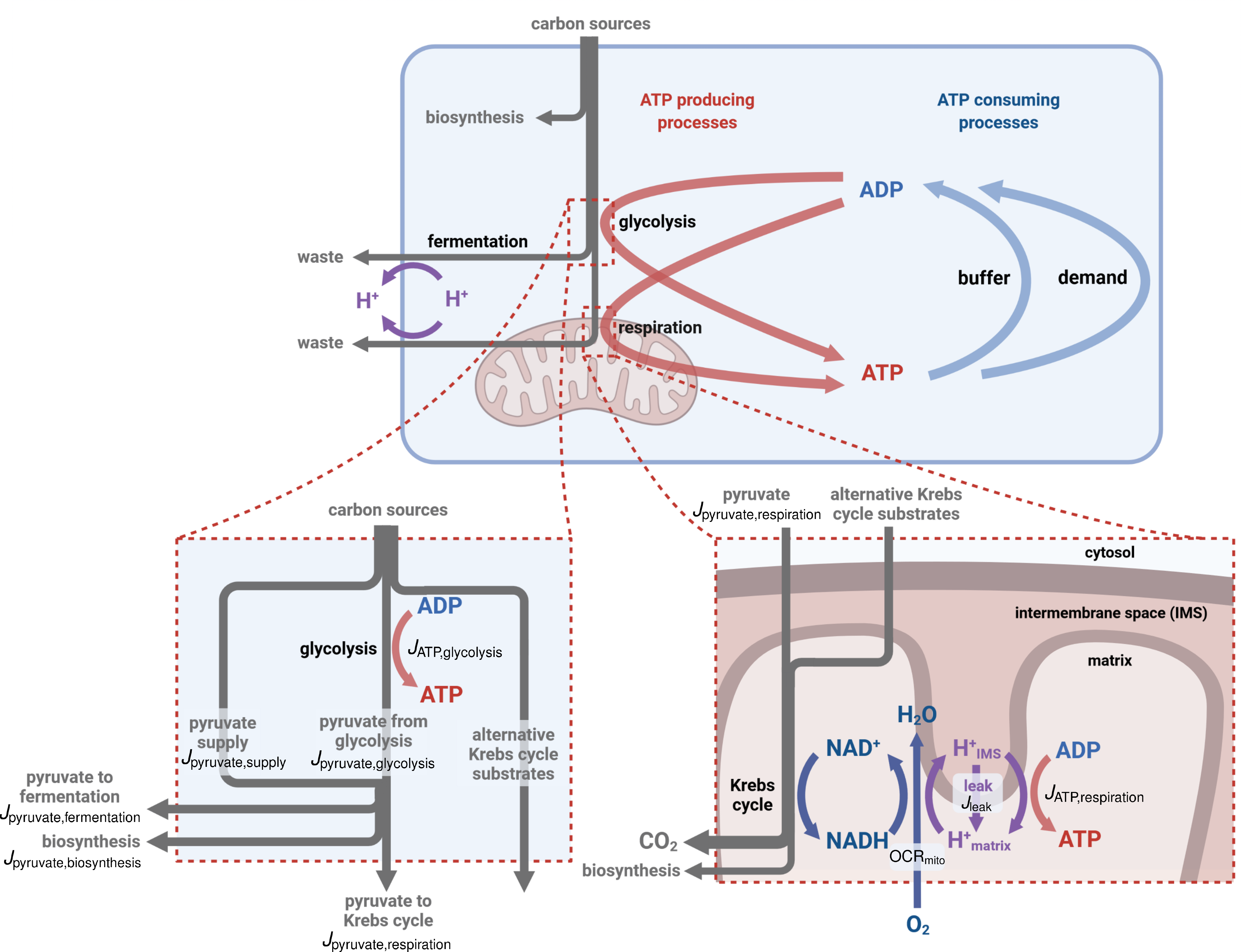}
    \caption{\textbf{Coupling between ATP production/consumption and other intra/extracellular fluxes.} \underline{Top left:} Glycolysis produces ATP through substrate-level phosphorylation, and breaks down substrates into species which serve as inputs for fermentation and respiration. Release of fermentation waste products is sometimes coupled to proton export, which leads to extracellular acidification. Respiration converts carbon inputs and oxygen to carbon dioxide, which is released from the cell and also contributes to extracellular acidification. \underline{Bottom left:} Pyruvate produced by glycolysis (and/or directly supplied) is consumed by fermentation and the Krebs cycle (respiration), or can be diverted to biosynthesis. \underline{Bottom right:} coupled cycles within mitochondria transduce free energy from breakdown of carbon skeletons in the Krebs cycle to power ATP synthesis. Reducing equivalents from glycolysis shuttled into the mitochondria are not shown here. \underline{Top right:} Within the framework of cellular activity modulation experiments, ATP-consuming processes can be conceptually separated into those sensitive to changes in ATP availability (``buffer'') and those which are insensitive (``demand''). Note that the ATP flux of some buffer processes may switch sign, in which case they become secondary ATP-producing processes. Portions of this figure were created with BioRender \cite{biorender}.}
    \label{fig:carbon_and_mito_detail}
\end{figure}

In many systems, ATP is produced primarily through glycolysis and respiration, which are catabolic pathways that transduce free energy from the breakdown of carbon sources to phosphorylate ADP. The fluxes through these pathways are coupled primarily through pools of carbon-containing metabolites and the electron carrier \ce{NAD(H)}. Glycolysis and respiration each directly produce ATP. Additionally, glycolysis reduces the electron carrier \ce{NAD+} to produce \ce{NADH}, as well as the carbon-containing compound pyruvate. These pyruvate molecules then serve as input to respiration, or another pathway, fermentation. Fermentation converts pyruvate to waste products without producing ATP, and serves to regenerate the cofactor \ce{NAD+}. In contrast, respiration involves oxidation of pyruvate to \ce{CO2} through the Krebs cycle, which is coupled to the electron transport chain (ETC) to produce ATP.
Fig. \ref{fig:carbon_and_mito_detail} illustrates the high-level couplings between ATP synthesis and hydrolysis to carbon and other fluxes. The relationships between these fluxes suggest which experimentally accessible reaction rates can be measured to infer ATP production rates.

\subsubsection{Glycolytic ATP production}

Glycolysis is a highly conserved ten-reaction pathway that converts glucose to pyruvate, which is then processed by fermentation or respiration. In the first portion of glycolysis, ATP is consumed to split glucose into two carbon skeletons that are prepared for the second, energy-yielding phase. In this phase each of the two skeletons is converted to a pyruvate molecule, producing one molecule of the electron carrier NADH and two molecules of ATP each. Thus, there is a net gain of 2 molecules of ATP and 2 molecules of NADH per input glucose that passes through the entirety of glycolysis \cite{alberts_2008}.

Other carbon sources, such as fructose and sucrose, can also enter glycolysis at intermediate stages. Additionally, a number of glycolytic intermediates are siphoned off for various biosynthetic reactions, including production of nucleotides, amino acids, and lipids. The relative magnitudes of biosynthetic and glycolytic fluxes can vary widely from system to system, resulting in different fractions of glycolytic substrates ultimately participating in the ATP-producing steps of glycolysis and being transformed into pyruvate \cite{frick2005characterization,mitsuishi2012nrf2}. These complications can be coarse-grained into an effective factor $ \alpha_\text{PPyr} $ which describes the average molar ratio of ATP directly produced by glycolysis to pyruvate produced by glycolysis, which can be measured in a given system using metabolic flux analysis (MFA) or other methods:
\begin{align}
    \begin{split}
        J_\text{ATP, glycolysis} &= \alpha_\text{PPyr} J_\text{pyruvate, glycolysis}
    \end{split}
    \label{eq:JatpJpyruvate}
\end{align}

We can then back out $J_\text{pyruvate, glycolysis}$ by considering how glycolytic, respiratory, fermentative, and biosynthetic fluxes are coupled through pyruvate. Pyruvate can be fermented (with flux $ J_\text{pyruvate, fermentation} $) to species like lactate, acetate, or ethanol depending on the specific system and nutrient conditions, oxidizing NADH back to \ce{NAD+} in the process; the fermentation products are then eliminated by the cell. Pyruvate can also be fed into the Krebs cycle and thus contribute to respiration (with flux $ J_\text{pyruvate, respiration} $), and can also be diverted to biosynthesis (with flux $J_\text{pyruvate, biosynthesis}$) \cite{fraenkel2011yeast}. Pyruvate may also be directly supplied in some systems (with flux $ J_\text{pyruvate, supply} $). At steady state, the flux balance on pyruvate is:
\begin{align}
    \begin{split}
        J_\text{pyruvate, glycolysis} &= J_\text{pyruvate, respiration} + J_\text{pyruvate, fermentation} + J_\text{pyruvate, biosynthesis} - J_\text{pyruvate, supply}
    \end{split}
    \label{eq:pyruvatebalance}
\end{align}
\noindent as illustrated in Fig. \ref{fig:carbon_and_mito_detail} (lower left detail). Under certain nutrient conditions, fermentation operates in reverse (e.g. conversion of lactate or ethanol to pyruvate), in which case $ J_\text{pyruvate, fermentation} $ is negative. $ J_\text{pyruvate, fermentation} $ and $ J_\text{pyruvate, supply} $ are typically determined by measuring cellular imports and exports, while $ J_\text{pyruvate, biosynthesis} $ can be determined by measurements of cellular composition and growth rate. Measuring $ J_\text{pyruvate, respiration} $, as well as characterizing the indirect contribution of glycolytic NADH production to respiratory ATP production, requires measurement of mitochondrial respiratory activity.

\subsubsection{Mitochondrial respiratory ATP production}
\label{sec:respATPprod}

Mitochondrial respiration consists of a series of free energy transduction cycles that link catabolic reactions to ATP production \cite{salway2016metabolism} (Fig. \ref{fig:carbon_and_mito_detail}, lower right detail). Free energy from breakdown of carbon sources, including pyruvate, carbohydrates, fatty acids, and amino acids is transduced to electron carriers NADH/\ce{NAD+} or \ce{FADH2}/\ce{FAD+}. The reducing equivalents supplied by these carriers are used by the electron transport chain (ETC) to reduce molecular oxygen and establish a proton gradient across the mitochondrial inner membrane. Finally, the free energy associated with this proton gradient is used to power ATP synthesis; this is achieved by proton flow through ATP synthase, a cross-membrane rotary motor. However, some fraction of protons leak directly through the lipid bilayer without passing through ATP synthase, dissipating free energy into heat.

At steady state, the net fluxes through the \ce{NADH}/\ce{NAD+} redox cycle, the \ce{H+} translocation cycle, and the ATP/ADP cycle are balanced with each other. Because the ETC consumes oxygen to oxidize NADH, mitochondrial oxygen consumption rate ($\text{OCR}_\text{mito}$) is proportional to the NADH oxidation flux (assuming \ce{FADH2} flux is negligible or proportional to NADH flux), and is proportional to the rate of proton extrusion from the mitochondrial matrix into the intermembrane space. The rate of proton extrusion, less the rate of proton leak through the mitochondrial inner membrane ($J_\text{leak}$), is the rate at which protons return to the matrix through ATP synthase. The rate of ATP synthesis by this process is thus given by:
\begin{align}
    J_\text{ATP,respiration} &= \alpha_\text{PO} \text{OCR}_\text{mito} - \alpha_\text{PO} \alpha_\text{HO}^{-1} J_\text{leak}
    \label{eq:jatpoxphos}
\end{align}

\noindent where fluxes of different species are related to each other by the \ce{H+}/O ratio, $\alpha_\text{HO}$ (the number of protons translocated per oxygen atom reduced) and the P/O ratio, $\alpha_\text{PO}$ (the number of ATP molecules synthesized per oxygen atom reduced in the absence of proton leak). $\alpha_\text{HO}$ is 10 for NADH oxidation and 6 for \ce{FADH2} oxidation. The exact value of $\alpha_\text{PO}$ is variable, but a typical value for mammalian cells is 2.5 for mitochondria using NADH and 1.5 when using \ce{FADH2} \cite{hinkle2005p,hinkle1991,brand1995}. These ratios are determined by the reaction and subunit stoichiometry of the enzymes catalyzing the reactions in mitochondria: $\alpha_\text{HO}$ is determined by complex I, III and IV of the electron transport chain; $\alpha_\text{PO}$ is determined by $\alpha_\text{HO}$, properties of ATP synthase, and transport of ATP from mitochondria to the cytoplasm by the enzyme adenine nucleotide translocase. Estimating $ J_\text{leak} $ is challenging, as we discuss in Section \ref{sec:complications}.

\subsubsection{Measuring combined ATP production by glycolysis and respiration}
\label{sec:integ_glyc_resp}

The NADH reducing equivalents produced by glycolysis can effectively be transported into mitochondria by shuttling mechanisms (e.g. the malate-aspartate shuttle, glycerol-3-phosphate shuttle) \cite{mookerjee2017quantifying}. This indirect contribution of glycolysis to total cellular ATP production is captured in the measurement of respiratory ATP production (Eq. \ref{eq:jatpoxphos}).

Pyruvate produced by glycolysis is not necessarily the only input to the Krebs cycle, and therefore it is not the only source of reducing equivalents for the electron transport chain in many systems \cite{martinez2020mitochondrial,filipp2012glutamine}. Instead, the flux of pyruvate from glycolysis to respiration is given by:
\begin{align}
    \begin{split}
        J_\text{pyruvate, respiration} &= \alpha_{\text{O}\text{Pyr}_{\text{glyc}}}^{-1} \text{OCR}_\text{mito}
    \end{split}
    \label{eq:Jpyrresp}
\end{align}
\noindent where $ \alpha_{\text{O}\text{Pyr}_{\text{glyc}}} $ is an empirical proportionality constant that is the ratio of mitochondrial OCR to the rate at which pyruvate is supplied to respiration by glycolysis. $ \alpha_{\text{O}\text{Pyr}_{\text{glyc}}} $ captures two variable factors: (1) the fraction of pyruvate produced by glycolysis, as opposed to pyruvate
supplied by other means, which is fed into the Krebs cycle, and (2) the variable contribution of pyruvate to the production of reducing equivalents (e.g. NADH). Krebs cycle intermediates can be siphoned off to participate in biosynthetic reactions, and some species can enter the cycle at a point other than Acetyl-CoA; $\alpha$-ketoglutarate is an example. Additionally, some organisms employ variants of the Krebs cycle (e.g. the glyoxylate shunt) which involve different reactions \cite{fraenkel2011yeast}; these differences are similarly reflected in $ \alpha_{\text{O}\text{Pyr}_{\text{glyc}}} $.

Combining Eqs. \ref{eq:JatpJpyruvate}-\ref{eq:Jpyrresp} we see that the total rate of ATP production by glycolysis and respiration is:
\begin{align}
    \begin{split}
        J_\text{ATP production} &= \alpha_\text{PPyr} \bigg( \alpha_{\text{O}\text{Pyr}_{\text{glyc}}}^{-1} \text{OCR}_\text{mito} + J_\text{pyruvate, fermentation}+ J_\text{pyruvate, biosynthesis} - J_\text{pyruvate, supply} \bigg) \\
        &\qquad+ \alpha_\text{PO} \text{OCR}_\text{mito} - \alpha_\text{PO} \alpha_\text{HO}^{-1} J_\text{leak}
    \end{split}
    \label{eq:ATPprod_integrated}
\end{align}

\noindent Eq. \ref{eq:ATPprod_integrated} provides a strategy to determine the flux of ATP production from glycolysis and respiration (which, at steady-state is equal to the flux of ATP consumption) by measuring mitochondrial OCR, fluxes associated with nutrient uptake, nutrient secretion, and biomass synthesis, and the flux associated with proton leak. Such an approach has been used to infer ATP production rates in mammalian cells by simultaneous measurement of both fermentation and respiration rates \cite{mookerjee2017quantifying,ferrick2008}. Alternative methods to measure $ J_\text{ATP, glycolysis} $ that do not rely on characterizing the pyruvate balance (e.g. measurements of glycolytic substrate uptake and biosynthesis rates) may be more suitable in other systems.

\subsection{Measuring glycolytic and respiratory fluxes using calorimetry
\label{sec:calorimetry}}

We shall now show that calorimetry provides an alternative means to determine the flux through respiration and glycolysis. In a constant-pressure calorimeter, the heat output measured by the calorimeter, $\dot{Q}$, equals the rate of of change of the enthalpy of the entire calorimetry sample, $H$; that is, $ \dot{Q} = \frac{dH}{dt}$, where a negative value of $\dot{Q}$ indicates the sample is producing heat. The enthalpy of the entire calorimetry sample, $H$, is the sum of the enthalpy of the biological sample, $H_\text{cells}$, and the enthalpy of the media that surrounds it, $ H_\text{extracellular} $. In cell biological systems, changes in enthalpy primarily result from chemical reactions converting one molecular species into another, in which case:

\begin{align}
    \dot{Q} &= \frac{dH_\text{cells}}{dt} + \frac{dH_\text{extracellular}}{dt} = \sum_{\text{species } i}  h_i \left( \dfrac{dn_{i,\text{cells}}}{dt} \right) + \sum_{\text{species } i}  h_i \left( \dfrac{dn_{i,\text{extracellular}}}{dt} \right)
    \label{eq:firstlaw4}
\end{align}

\noindent where $n_{i,\text{cells}}$ and $n_{i,\text{extracellular}}$ are the number of moles of species $i$ within the biological sample and the surrounding media, respectively, and $h_i$ is the molar enthalpy of that species. In the special case that cells are at steady state, i.e. they are not growing and their composition is not changing in time, $\frac{dn_{i,\text{cells}}}{dt} = 0$. Then, the change in the concentrations of species $i$ in the surrounding media is equal to the difference in cellular import flux, $J_{i,\text{import}}$, and export flux, $J_{i,\text{export}}$, of that species. Thus, Eq. \ref{eq:firstlaw4} becomes:

\begin{align}
    \dot{Q}_\text{ss} &= \sum_{\text{species } i}  h_i J_{i,\text{import}} - \sum_{\text{species } i}  h_i J_{i,\text{export}} = \sum_{\text{rxns } k} \Delta h_{\cdot,k} J_{\cdot,k}
    \label{eq:firstlaw7}
\end{align}

\noindent where $ J_{\cdot,k} $ is the flux of an index species (denoted by $ \cdot $) through the net reaction $ k $, which proceeds from cellular imports to cellular exports, and $ \Delta h_{\cdot,k} $ is the enthalpy of reaction per mole of the index species. Eq. \ref{eq:firstlaw7} reveals that in steady state the heat production measured by calorimetry $\dot{Q}_\text{ss}$ reflects the difference in enthalpy between the cellular imports and waste, i.e. the net reaction of the cell. This relationship forms the basis for calorimetric measurements of net metabolic fluxes.

\begin{figure}[!htpb]
    \centering
    \includegraphics[width=\textwidth]{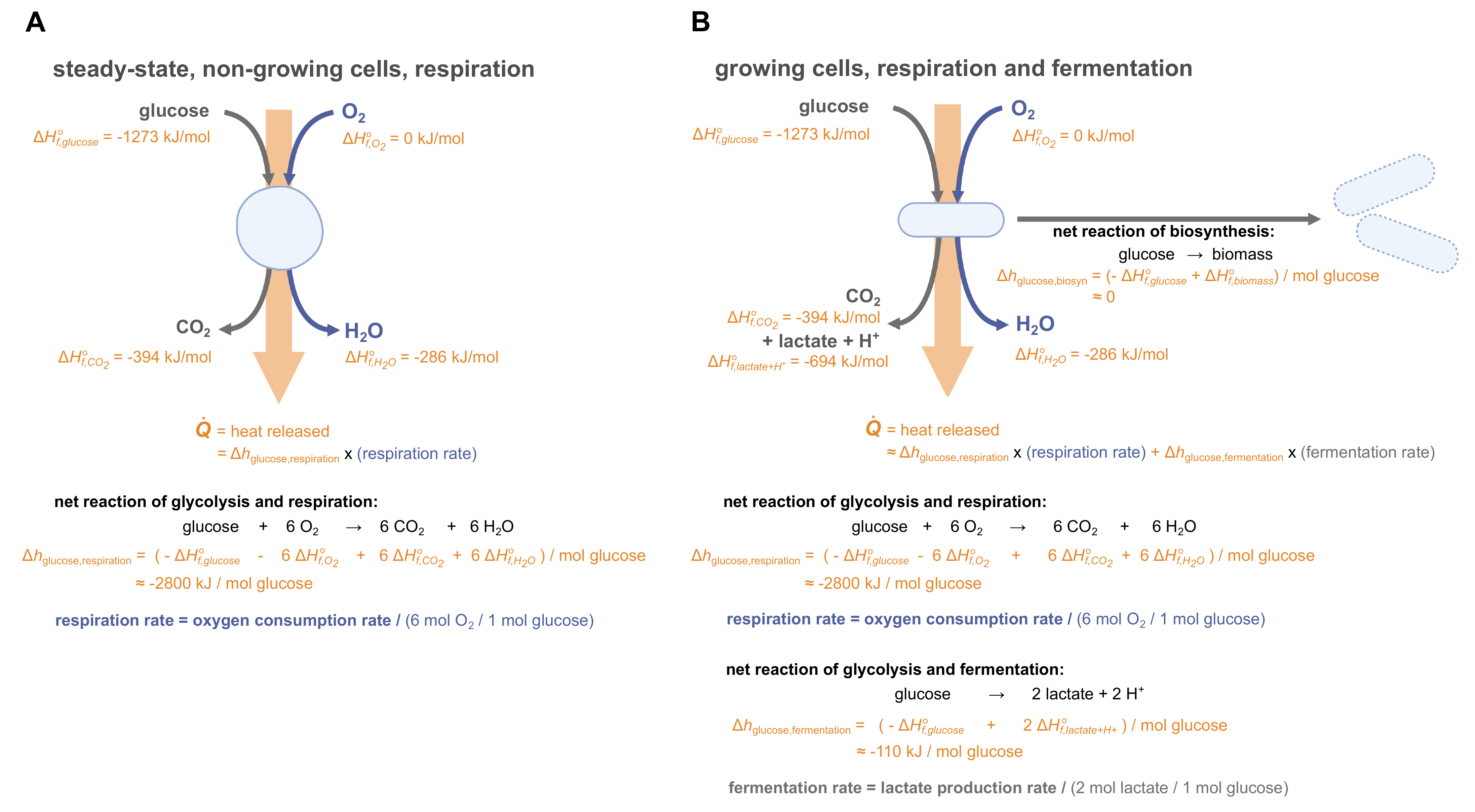}
    \caption{\textbf{Coupling between chemical and enthalpy fluxes results in a straightforward relationship between carbon source and oxygen consumption rates and heat production.} A: Net reaction of steady-state, non-growing cells that consume glucose and respire it completely. The enthalpy production is proportional to oxygen consumption rate. B: Net reaction of growing cells that consume glucose and ultimately perform a combination of respiration and fermentation to lactate. The enthalpy production depends on the relative fluxes through fermentation and respiration, but does not depend strongly on biomass production. Portions of this figure were created with BioRender \cite{biorender}.}
    \label{fig:enthalpy}
\end{figure}

For steady-state cells that respire all of the carbon that they import, e.g. cell-cycle arrested oocytes, there is a direct relationship between respiratory flux and heat production (Fig. \ref{fig:enthalpy}A). Consider a steady-state cell that imports glucose and oxidizes it completely to carbon dioxide and water. The net reaction of such a cell is \ce{C6H12O6 + 6O2 -> 6 CO2 + 6H2O}. Because glucose, carbon dioxide, and water have enthalpies of formation of -1273, -394, and -286 kJ/mol, respectively, the total enthalpy of reaction is $\Delta h_\text{glucose,respiration} \approx$ -2800 kJ/(mol glucose consumed). The heat output for such a system is
$\dot{Q}_\text{ss} = \Delta h_\text{glucose,respiration}  J_\text{glucose,import} = \frac{\Delta h_\text{glucose,respiration}}{6} \text{OCR}_\text{mito}$ where $J_\text{glucose,import}$ is the rate of glucose consumption. Thus, approximately 467 kJ are liberated for every mole of \ce{O2} consumed, and the heat output measured is straightforwardly related to the OCR measurements discussed in Sec. \ref{sec:respATPprod}. While this calculation is for the specific case of purely respiratory metabolism using glucose as the substrate, there is a long-standing empirical observation known as Thornton's rule \cite{thornton_xv_1917} that the heat evolved from consuming a given amount of oxygen, called the oxycaloric coefficient, is approximately constant ($\sim$-455 kJ/mol \ce{O2}) across different substrates. Thornton's rule enables inference of OCR from calorimetry measurements without information about the identity of the respiratory substrate. This suggests an approach to infer ATP production flux from calorimetry: heat flux may be converted to an equivalent OCR using Thornton's rule, and OCR can then be converted to the corresponding ATP production flux using Eq. \ref{eq:jatpoxphos} and measured values of the proton leak flux and stoichiometric ratios.
 
To extend this analysis beyond purely respiratory metabolism, let us next consider steady-state cells that obtain all their energy from anaerobic glycolysis. A common scenario involves import of glucose and fermentation to lactate, in which case calorimetry measures the enthalpy flux associated with the reaction \ce{C6H12O6 -> 2 C3H5O3- + 2H+}. The enthalpy of this reaction is $\Delta h_\text{glucose,fermentation} \approx$ -110 kJ/(mol glucose consumed); this is far smaller than $\Delta h_\text{glucose,respiration} \approx$ -2800 kJ/(mol glucose consumed) because fermentation results in the incomplete oxidation of glucose. Using the enthalpy of reaction, one can convert the heat flux to a glycolytic flux, which can in turn be converted to ATP production flux using Eq. \ref{eq:JatpJpyruvate} if the effective stoichiometry is known.
 
For non-growing, steady state cells which perform respiration, glycolysis and fermentation, the heat output measured by calorimetry will reflect the fluxes through all of these reactions. Consider a cell which consumes glucose, some of which is completely respired, with the rest fermented to lactate. The heat output for such a system is $\dot{Q}_\text{ss} = \sum_{\text{rxns } k} \Delta h_{\cdot,k} J_{\cdot,k} = \frac{\Delta h_\text{glucose,respiration}}{6}   \text{OCR}_\text{mito} + \frac{\Delta h_\text{glucose,fermentation}}{2}   J_\text{lactate,export} \approx 467 \text{kJ/(mol } \ce{O2} ) \text{ OCR}_\text{mito}  + 54 \text{kJ/(mol lactate) } J_\text{lactate,export}$, where $J_\text{lactate,export}$ is the rate of lactate secretion (i.e. the rate of fermentation). Thus, if heat flux and OCR are measured, the rate of fermentation can be inferred (or if heat flux and fermentation are measured, OCR can be inferred). More generally, the heat associated with respiration can be calculated from Thornton's rule, while the heat associated with fermentation will depend on the particular fermentation products. The ratio between the predicted respiration-associated heat output and that measured by calorimetry has been used to probe how the mode of metabolism changes over time in various systems \cite{hansen_use_2004,kemp_calorimetric_1991,lerchner_direct_2019}. 

We have thus far only considered cells at steady state, but many systems of interest are not in steady state, in which case the internal energy of the cells is constantly changing. For growing cells, biomass increases over time, and thus $\frac{dn_{i,\text{cells}}}{dt} \neq 0$; Eq. \ref{eq:firstlaw4} can then be written as:
 
\begin{align}
    \dot{Q} &= \sum_{\text{net biosyn rxns }k}  \Delta h_{\cdot,k} J_{\cdot,k} + \sum_{\text{net non-biosyn rxns }k'}  \Delta h_{\cdot,k'} J_{\cdot,k'}
    \label{eq:firstlaw8}
\end{align}

\noindent where $k$ indexes net biosynthetic reactions, from imported compounds to biomass components, e.g. glucose to lipids or glucose to protein; $ J_{\cdot,k} $ represents the flux of an index species (denoted by $ \cdot $) through reaction $ k $; and $ \Delta h_{\cdot,k} $ is the enthalpy of that reaction per mole of the index species. Similarly, $k'$ indexes net non-biosynthetic reactions (including ATP-producing pathways); $ J_{\cdot,k'} $ represents the flux of an index species (denoted by $ \cdot $) through reaction $ k' $; and $ \Delta h_{\cdot,k'} $ is the enthalpy of that reaction per mole of the index species. The two dominant net non-biosynthetic reactions that we consider are 1) respiration of imported carbon sources, and 2) conversion of imported carbon sources to fermentation products. In cells consuming glucose and performing glycolysis, respiration, and fermentation, these correspond to glycolysis followed by respiration and glycolysis followed by fermentation, respectively.
 
However, the enthalpy of formation of biomass is very close to that of typical carbon sources for cells. Experimental measurements in \textit{E. coli} indicate that the enthalpy of net anabolic reactions is very near zero \cite{battley_enthalpy_1992}, which simplifies equation \ref{eq:firstlaw8} to:

\begin{align}
    \begin{split}
        \dot{Q} &\approx \sum_{\text{non-biosyn rxns }k'}  \Delta h_{\cdot,k'} J_{\cdot,k'} \approx \Delta h_{\ce{O2},\text{respiration}} \text{OCR}_\text{mito} + \Delta h_{\text{ferm prod,fermentation}} J_{\text{ferm prod,export}}.
        \label{eq:firstlaw9}
    \end{split}
\end{align}

\noindent where $\Delta h_{\ce{O2},\text{respiration}} $ is the enthalpy associated with the net reaction of respiration from import of carbon source to export of \ce{CO2}, per mole of oxygen consumed (which is $\sim$-455 kJ/mol \ce{O2} by Thornton's rule); $ \Delta h_{\text{ferm prod,fermentation}} $ is the enthalpy associated with the net reaction of import of carbon source to export of fermentation products, per mole of fermentation product exported; and $ J_{\text{ferm prod,export}} $ is the rate at which said fermentation product is exported from the cell.

In summary, changes in biomass often contribute negligibly to the heat flux measured by calorimetry, which can therefore be related to flux through glycolysis, fermentation, and respiration even in growing cells (Fig. \ref{fig:enthalpy}B). Thus, calorimetry reflects metabolic fluxes even in non-steady-state biological systems, such as developing \textit{Drosophila} and zebrafish embryos \cite{song_energy_2019,rodenfels_contribution_2020}.

\section{Measuring energetic costs by process activity modulation}
\label{sec:complications}

Equation \ref{eq:fluxes1c} illustrates how ATP production and consumption are linked; in a system approximately at steady state, these two fluxes are balanced. Therefore, as described above, an intuitively appealing strategy for inferring the energetic cost of a process is to change the activity level of the process and measure concomitant changes in energy production rate.

Several authors have inhibited processes such as translation and measured changes in oxygen consumption rate \cite{wieser2001hierarchies,smith1995protein}, though inferring the energetic cost of a process from a single activity modulation experiment presupposes that the perturbation leaves the rates of other processes unchanged. Practically, such separability of different processes is unlikely; the activity of different processes can be linked. For example, inhibiting transcription has downstream effects on translation; inhibiting translation prevents production of enzymes whose operation involves ATP hydrolysis; and the reduced activity of these ATPases can change the cellular environment, such as through pH changes which affect a range of proteins. Each level of these indirect effects affects additional cellular machinery, and therefore has implications for ATP consumption. Thus, the observed changes in metabolic fluxes are a superposition of the direct effects of the applied perturbation, as well as indirect first order effects, second order effects, and so on. The downstream effects of perturbations can manifest on multiple timescales. For example, changes in metabolite concentrations and signaling may occur rapidly, while changes in protein expression take place more slowly \cite{buchenberg_time-resolved_2017,maier_quantification_2011,Kresnowati_when_2006}. Thus, extracting information about energetic costs from activity modulation experiments requires a systematic accounting of the effects of a perturbation on other cellular machinery.

In the case of a steady-state system, ATP production is balanced by ATP consumption. In this case, if the activity of a cellular process $p$ is altered, changing the ATP demand of that process $J_{\text{ATP},p}$ by some amount $ \Delta J_{\text{ATP},p}$, and the system reaches a new steady state, the difference between the ATP balance before and after inhibition indicates the cost associated with the difference in the activity of that process. Combining Eq. \ref{eq:fluxes1c} (at steady-state) and Eq. \ref{eq:ATPprod_integrated}, we obtain the change in the ATP consumption flux of process $p$:

\begin{align}
    \begin{split}
        \Delta J_{\text{ATP},p} &= \Delta J_\text{ATP,glycolysis} + \alpha_\text{PO} \Delta \text{OCR}_\text{mito} - \alpha_\text{PO} \alpha_\text{HO}^{-1} \Delta J_\text{leak} - \sum_{k' \neq p} \Delta J_{\text{ATP demand},k'} - \sum_{k'' \neq p} \Delta J_{\text{ATP buffer},k''}
    \end{split}
    \label{eq:ATPcost_inhibitionexpt}
\end{align}

\noindent If the process is completely inhibited, $\Delta J_{\text{ATP},p} = J_{\text{ATP},p}$ is the total energetic cost of the process. Eq. \ref{eq:ATPcost_inhibitionexpt} provides a quantitative framework to measure energetic costs from activity modulation experiments. This equation connects measurements of experimentally accessible fluxes to the energetic costs of interest.

In the simplest possible case, when inhibition of the target process does not impact any other ATP-consuming process or mitochondrial leak, then 
$ \Delta J_{\text{leak}}$, $ \Delta J_{\text{demand}, k' \neq p}$, and $ \Delta J_{\text{buffer}, k'' \neq p}$ are zero, and measuring the change in ATP production by glycolysis and respiration will yield the energetic cost of the inhibited process. However, in general $ J_{\text{leak}}$, $ J_{\text{demand}, k' \neq p}$, $ J_{\text{buffer}, k'' \neq p}$, and other coupled cellular processes could change as a result of modulating the activity level of a process of interest. Therefore, to determine the energetic cost of process $ p $ using an activity modulation experiment, one must either (1) directly measure the change in individual fluxes on the right hand side of Eq. \ref{eq:ATPcost_inhibitionexpt} in each experiment, which is generally difficult, or (2) develop mechanistic models of flux regulation that capture the generic response of each flux term to changes in experimentally accessible metabolite concentrations or chemical potentials (e.g. ATP concentration, ATP/ADP ratio, proton motive force, NADH/\ce{NAD+} ratio) that can be more easily measured in individual activity modulation experiments.

Below we highlight key challenges associated with measuring each of the flux terms on the right hand side of Eq. \ref{eq:ATPcost_inhibitionexpt} to determine energetic costs from activity modulation experiments:

\begin{itemize}
    
    \item \underline{\textit{Measuring} $\Delta J_\text{ATP,glycolysis}$ \textit{requires characterization of flux partitioning}.} There is not necessarily a constant ratio between the glucose routed through glycolysis and anabolic pathways under different conditions \cite{frick2005characterization,mitsuishi2012nrf2}. Because glycolytic intermediates are siphoned off for biosynthesis, the flux through a single step of the pathway does not directly indicate ATP payoff from glycolysis. Similarly, inferring glycolytic pyruvate input to the Krebs cycle is challenging because of the many branch points within the Krebs cycle which can introduce or siphon off intermediates. These complications are captured in the effective stoichiometric coefficients $\alpha_\text{PPyr}$ and $ \alpha_{\text{O}\text{Pyr}_{\text{glyc}}}$. Measuring these quantities may require high-resolution mapping of metabolic fluxes, as discussed in Sec. \ref{sec:methods}.
    
    \item \underline{$\alpha_\text{HO}$ \textit{ and } $\alpha_\text{PO}$ \textit{are substrate-dependent.}} For substrates whose metabolism involves NADH (e.g. pyruvate) $\alpha_\text{HO}$ is 10 and $\alpha_\text{PO}$ is typically 2.5 in mammalian cells, but for those whose metabolism involves \ce{FADH2} (e.g. succinate) $\alpha_\text{HO}$ is 6 and $\alpha_\text{PO}$ is typically 1.5 \cite{hinkle2005p,hinkle1991,brand1995}. The relative fluxes of the various substrates which enter the Krebs cycle at different stages must be determined in order to estimate the net $\alpha_\text{HO}$ and $\alpha_\text{PO}$.
    
    \item \underline{$\Delta J_\text{leak}$ \textit{depends on proton motive force}}. Proton leak represents the fraction of mitochondrial OCR that does not contribute to ATP synthesis due to the permeability of the mitochondrial inner membrane to protons and the possible presence of uncouplers. This flux can be measured by OCR following inhibition of ATP synthase activity. Proton leak may not necessarily remain constant after perturbations: process activity modulation may change ATP synthesis rate, which could in turn alter the proton gradient across the mitochondrial inner membrane and thus the proton leak flux. Combining measurements of proton motive force (a combination of the proton gradient and the membrane potential) with a model of how leak flux depends on the proton motive force enables estimation of $\Delta J_\text{leak}$ \cite{divakaruni2011}.
    
    \item \underline{\textit{Off-target effects and coupling between processes may lead to non-vanishing} $\Delta J_\text{demand}$.}
    Activities of cellular processes of interest are often altered through pharmaceutical perturbations which may have off-target effects; for example, cycloheximide, an inhibitor of protein synthesis, is thought to directly affect metabolism at high concentrations \cite{wieser2001hierarchies,ellis1970specificity}. Moreover, different ATP-demanding processes in the cell can be coupled, and their values could each change upon perturbation, either as a direct result of the perturbation or due to indirect effects. For example, the product of one ATP (equivalent)-consuming process could be required for a different ATP-consuming process, as in the case of transcription and translation. Gross changes in cellular physiology caused by a perturbation (e.g. pH, osmolarity) could also affect many different demand and buffering processes.
    
    \item \underline{\textit{Measuring} $\Delta J_\text{buffer}$ \textit{requires characterization of ATP flux-buffering mechanisms.}} ATP-dependent processes can interact through the shared ATP/ADP pool: changes to this pool caused by perturbing one ATP-dependent process could alter ATP/ADP concentrations, affecting other machinery sensitive to ATP levels in turn. Because $J_\text{buffer}$ could compensate for flux changes induced by process activity modulation, it must be estimated by identifying individual buffering enzymes and characterizing their sensitivity to changes in ATP/ADP concentrations \cite{buttgereit1995hierarchy,brown1992control}.

\end{itemize}

In addition to the challenges discussed above, we note that the applicability of Eq. \ref{eq:ATPcost_inhibitionexpt} is limited to organisms producing ATP exclusively from glycolysis and oxygen-dependent respiration, which is the case for many systems of interest. However, some organisms may utilize different modes of energy metabolism, and Eq. \ref{eq:ATPcost_inhibitionexpt} can be generalized accordingly by including the corresponding fluxes. For example, oxygen is not the only electron acceptor in respiration. Many bacteria perform respiration with terminal electron acceptors such as fumarate and nitrate instead of oxygen \cite{jacobs1976}. Recent work has shown that under conditions of hypoxia or inhibition of Complex IV of the mitochondrial electron transport chain, mammalian cells can similarly use fumarate as the terminal electron acceptor \cite{spinelli2021}. Under these conditions OCR does not quantify respiratory activity, and alternative methods such as the measurement of NADH/\ce{NAD+} flux \cite{Yang2021elife} may be necessary. Additionally, alternative ATP production pathways, such as photosynthesis or formation of ATP from AMP and pyrophosphate represent major sources of ATP in some systems \cite{allen2002,james2016pyrophosphate}. Moreover, Eq. \ref{eq:ATPcost_inhibitionexpt} only considers ATP production and consumption, but other metabolites also provide free energy to power cellular activities. For example, NADPH plays a central role in oxidative stress response activities and in biosynthetic reactions, and its turnover rate constitutes a significant portion of the cost of biosynthesis \cite{salway2016metabolism,wu2011beneficial}. Measurements of fluxes of NADPH and other metabolites will be necessary to provide a complete picture of the energetic cost of cellular processes.

In summary, the most direct means to determine the energetic cost of a cellular process is to measure how altering the activity of that process impacts ATP fluxes. Eq. \ref{eq:ATPcost_inhibitionexpt} provides a systematic means to interpret such experiments. This equation enumerates the relevant, individual fluxes that will be present in many biological systems of interest. Each of these fluxes is challenging to charaterize, and some must be estimated by combining multiple types of measurements.

\section{Methods and technologies for measuring fluxes and concentrations}
\label{sec:methods}

Here we review a selection of the methods available to measure the metabolic fluxes referred to in Eq. \ref{eq:ATPcost_inhibitionexpt} and the associated metabolite concentrations. Technologies for flux and concentration measurements have matured considerably over the last two decades; modern metabolomics- and spectroscopy-based methods have enabled measurements with previously inaccessible spatial and temporal resolution \cite{law2022integrative}. These new technologies have produced a wealth of flux and metabolite concentration data in living cells, but it remains a challenge to integrate these data to infer the energetic cost of specific cellular processes. No single technique can accurately measure all the fluxes presented in Eq. \ref{eq:ATPcost_inhibitionexpt}.

\subsection{Measurements of respiration rate}
Oxygen is a common terminal electron acceptor in the mitochondrial electron transport chain (ETC), which drives respiratory ATP generation. Thus, oxygen consumption rate (OCR) is frequently used as a metric for global metabolic activity of aerobic organisms. In many types of eukaryotic cells, mitochondrial respiration accounts for the majority of OCR \cite{brand2011assessing}; hence, cellular OCR is a reliable proxy for mitochondrial ETC flux. For cells where non-mitochondrial OCR is substantial, mitochondrial inhibitors can be applied to separate the mitochondrial and non-mitochondrial OCR. In this section, we review techniques that measure cellular and mitochondrial OCR, which are summarized in Table \ref{tab:respiration}. 

Sealed-chamber respirometry is used to measure cellular OCR by determining the rate of decrease of oxygen level in an airtight chamber containing cells \cite{houghton1996,ferrick2008,gnaiger2000}. It is usually applied to a population of cells and can resolve OCR dynamics with minute-level time resolution. Open-chamber respirometry measures OCR in a half-open chamber by characterizing the steady-state spatial gradient in oxygen concentration across the chamber, which is established by the balance of oxygen diffusion from the open end and oxygen consumption of cells located at the other end \cite{lopez2005}. Establishing the concentration gradient may take a few hours, placing a lower bound on the time required to make the first measurement. The advantage of this technique is that the chamber can be engineered to resolve the OCR of a small number of cells, which is particularly useful when studying oocytes and embryos \cite{lopez2005}. 

Respirometry is a bulk technique that gives the net oxygen consumption of the entire sample. In contrast, fluorescence lifetime imaging microscopy (FLIM) has recently been established as a robust method to measure mitochondrial ETC flux with optical resolution, enabling the characterization of subcellular spatial variations \cite{Yang2021elife}. This is achieved through model-based inference using FLIM of mitochondrial NADH. This method requires mitochondrial respiration to be at steady or quasi-steady state, which holds as long as dynamics of interest are slower than the kinetic rates of mitochondrial enzymes, which are on the order of seconds. This technique can potentially resolve OCR dynamics on the order of seconds. 

\begin{table}[h]
\tabcolsep7.5pt
\caption{Common methods for measuring respiration rate}
\label{tab:respiration}
\begin{center}
\begin{tabular}{|m{0.30\textwidth}|m{0.18\textwidth}|m{0.07\textwidth}|m{0.08\textwidth}|m{0.27\textwidth}|}
\hline
Method & Timescale of measurement & Spatial resolution & Time series possible? & Examples \\ \hline
Sealed-chamber respirometry & minutes & none & yes & mouse embryos \cite{houghton1996}; tissue culture cells \cite{ferrick2008}; isolated mitochondria \cite{gnaiger2000} \\ \hline
Open-chamber respirometry & hours & none & yes & bovine oocytes \cite{lopez2005} \\ \hline
Fluorescence lifetime imaging microscopy (FLIM) & seconds & yes & yes & mouse oocytes and tissue culture cells \cite{Yang2021elife} \\ \hline
\end{tabular}
\end{center}
\end{table}

\subsection{Measurements of nutrient import and waste export}
\label{sec:methods_extracellular}

Extracellular measurements of nutrients and waste products are often used to infer the rates of glycolysis and fermentation. Each of these measurements requires that cells consume nutrients (e.g. glucose) and produce waste products (e.g. ethanol, acetate, lactate/\ce{H+}) at a quasi-steady rate. By analyzing the composition of the media, one can infer the rate of the pathway associated with the production or consumption of the analyte.
Methods for measuring nutrient import and waste export vary widely in their principles of operation and their limitations. The primary characteristics of several common methods are listed in Table \ref{tab:importexport}. 

Colorimetric and electrochemical assays employ enzyme-linked reactions with absorbance, fluorescence, or electrical signal-based readouts. These methods can be high throughput; they are often designed for parallel measurements in multi-well plates. However, the slow degradation of fluorescent compounds and enzymes can hamper accurate quantification.
Liquid chromatography (LC) involves separation of metabolites using chromatography, followed by quantification on the basis of absorbance by optical detectors, or mass-to-charge ratio (using mass spectrometry (MS)). LC-MS measurements are subject to complex caveats beyond the scope of this work, which have been described elsewhere \cite{lu2017metabolite}. Nuclear magnetic resonance (NMR) spectroscopy relies on the distinct frequency spectra of nuclei in different molecules to simultaneously quantify several nutrients/waste products in media. The number of species whose concentrations can be accurately resolved by NMR is frequently limited by spectral overlap of different compounds \cite{allen1997metabolite}.
Measurements of extracellular acidication rate (ECAR) are often performed in tandem with oxygen consumption rate measurements in commercial instruments like the Seahorse Extracellular Flux Analyzer (Agilent). Some cells perform lactic acid fermentation, producing lactate anions which are then exported by a lactate/\ce{H+} symport mechanism, acidifying media. Acidification is treated as a proxy for fermentation rate. However, respiratory \ce{CO2} production also contributes to acidification, and corrections must be applied to accurately infer fermentation rate \cite{mookerjee2017quantifying}. Additionally, this method assumes that all lactate produced is eliminated, which may not be the case in all systems \cite{hui2017glucose,schmidt2021ocr}.

The timescale of extracellular composition measurements is typically limited by how long it takes cells to significantly change the concentrations of nutrients or waste products in the medium. The length of a time course will vary greatly depending on the concentrations of nutrients or metabolites under consideration, and the fluxes involved in the turnover of their respective pools; rates can be calculated from serial samples of the same population.

\begin{table}[h]
\tabcolsep7.5pt
\caption{Common methods for measuring nutrient import and waste export}
\label{tab:importexport}
\begin{center}
\begin{tabular}{|m{0.30\textwidth}|m{0.18\textwidth}|m{0.07\textwidth}|m{0.08\textwidth}|m{0.27\textwidth}|}
\hline
Method & Timescale of measurement & Spatial resolution & Time series possible? & Examples \\ \hline
Colorimetric or electrochemical assays  & minutes to hours & none & no & yeast \cite{bagamery2020putative}; mouse \cite{bernier2020microglial} \\ \hline
Liquid chromatography (LC) with optical detection or mass spectrometry (MS) – media analysis & minutes to hours  & none & no & bacteria and yeast \cite{paczia2012extensive}; stem cells \cite{ferreiro2011automated} \\ \hline
Nuclear magnetic resonance (NMR) spectroscopy & minutes to hours & none & no & bacteria \cite{birkenstock2012exometabolome}; human bodily fluids \cite{duarte2014nmr}; rat brain \cite{koga1988measurement} \\ \hline
Extracellular Acidification Rate (ECAR) & minutes & none & yes & tissue culture cells \cite{mookerjee2017quantifying} \\ \hline
\end{tabular}
\end{center}
\end{table}

\subsection{Intracellular flux measurements}

Here, we consider intracellular fluxes to include those pathways whose rates are not solely determined by measurements of nutrient uptake, waste secretion, or oxygen consumption rate. Resolving these rates enables inference of flux partitioning at key branch points in metabolism, such as splits between bioenergetically relevant fluxes (e.g. glycolysis and Krebs cycle fluxes), and biosynthetic fluxes. Another example is the direct measurement of ATP synthesis flux, which is not constrained solely by respiration rate, but depends on proton leak as well.

Here we consider five examples of intracellular flux measurement methods: nuclear magnetic resonance (NMR) spectroscopy, \ce{^{31}P} saturation transfer magnetic resonance spectroscopy (MRS), metabolic flux analysis (MFA), kinetic flux profiling (KFP), and stimulated Raman spectroscopy (SRS), which each report different quantities: NMR can be used to probe the absolute values of net fluxes, MRS is used to estimate total cellular ATP production flux, MFA is used to measure ratios of net fluxes at branch points in metabolic pathways, KFP is used to measure the absolute values of gross fluxes, and SRS is used to measure the concentrations of specific chemical bonds. Their primary characteristics are summarized in Table \ref{tab:intracellular}.

\ce{^{31}P} MRS involves selective saturation of the $\gamma$-phosphate resonance of ATP in steady-state, followed by monitoring of changes in free phosphate signal over time to infer total cellular ATP synthesis rate \cite{brown197731p}. This method is nondestructive, but measurement times can be long in some cases. The accuracy of this method can in theory be affected by production of free cytosolic phosphate by de-phosphorylation of abundant intermediates (e.g. fructose-1,6-bisphosphate), which may lead to underestimation of ATP-to-free phosphate flux \cite{brindle1988phosphorous}. \ce{^{31}P} MRS is not widely used in modern bioenergetics studies.

For metabolic flux analysis (MFA), cells are supplied with labeled nutrients, and different branches in a metabolic pathway produce distinct steady-state labeling patterns in downstream metabolites. The ratiometric nature of MFA enhances the accuracy of flux partitioning estimates. However, MFA results must be constrained by extracellular flux measurements to infer absolute fluxes. The time required to achieve steady-state labeling depends on the system and pathway of interest, but can be short in some cases \cite{zamboni201113c,antoniewicz2015methods}. Recent work has focused on extending MFA to non-steady-state systems \cite{leighty2011dynamic,cheah2018isotopically}.

In kinetic flux profiling (KFP), a time course of isotope labeling patterns is measured by LC-MS following a rapid switch from an unlabeled to labeled nutrient. Assuming that metabolic reactions can be accurately modeled by first-order kinetics, rates of different reactions can be calculated. Labeling can be quite fast (e.g. $t_{1/2} < 5$ min for central metabolites in \textit{E. coli}), and thus KFP may be extended to non-steady-state systems \cite{yuan2008kinetic}.

Stimulated Raman spectroscopy (SRS) enables quantification of specific chemical bonds on the basis of their vibrational signature. By supplying labeled nutrients to cells and monitoring the concentration of specific bonds, one can observe their incorporation over time. However, in theory, both nutrient uptake and incorporation of bonds into biomass may contribute to the observed signal when using certain substrates; there may also be slight spectral shifts depending on the biomolecules into which the labeled bonds are incorporated \cite{zhang2019spectral}.

\begin{table}[h]
\tabcolsep7.5pt
\caption{Common methods for measuring intracellular fluxes}
\label{tab:intracellular}
\begin{center}

\begin{tabular}{|m{0.30\textwidth}|m{0.18\textwidth}|m{0.07\textwidth}|m{0.08\textwidth}|m{0.27\textwidth}|}
\hline
Method & Timescale of measurement & Spatial resolution & Time series possible? & Examples \\ \hline
Nuclear magnetic resonance (NMR) spectroscopy & minutes to hours & none & yes & human brain \cite{shen1999determination} \\ \hline
\ce{^{31}P} saturation transfer magnetic resonance spectroscopy (MRS) & minutes to hours & none & yes & yeast \cite{alger1982vivo}; rat kidney \cite{freeman1983energetics}; rat heart \cite{matthews1981steady} \\ \hline
Metabolic Flux Analysis (MFA) & hours & none & no & \textit{B. subtilis} \cite{fischer2005large}; \textit{E. coli} \cite{fischer2003metabolic}; yeast \cite{van2005metabolic} \\ \hline
Kinetic Flux Profiling (KFP) & minutes & none & yes & \textit{E. coli} \cite{yuan2008kinetic} \\ \hline
Stimulated Raman Spectroscopy (SRS) & minutes to hours & yes & yes & tissue culture cells \cite{hu2015vibrational,zhang2019spectral}, bacteria \cite{hong2018antibiotic} \\ \hline
\end{tabular}

\end{center}
\end{table}

\subsection{Intracellular metabolite concentration measurements}

Direct measurements of the fluxes on the right hand side of Eq. \ref{eq:ATPcost_inhibitionexpt} can be challenging. An alternative method to obtain these fluxes is to develop mechanistic or empirical models that relate fluxes such as OCR, proton leak and ATP buffering, to experimentally accessible quantities such as intracellular metabolite concentrations. Biophysical models of mitochondrial metabolic pathways are proving to be useful in revealing the relationships between fluxes and concentrations \cite{beard2005biophysical,korzeniewski2001model,Yang2021elife,jin2002kinetics,chang2011modeling}.

Fluorescent biosensors have recently enabled measurements of key metabolite concentrations in living cells with subcellular resolution. Examples include ATP concentration \cite{imamura2009visualization}, ATP/ADP ratio \cite{berg2009genetically}, mitochondrial membrane potential \cite{perry2011mitochondrial}, NADH/\ce{NAD+} ratio \cite{hung2011imaging}, pyruvate concentration \cite{san2014imaging}, lactate concentration \cite{san2013genetically} and glucose concentration \cite{diaz2019quantitative}. Sensors must be chosen so that their dynamical range is well-matched with the concentration of the analyte in the system of interest. It is also important to correct for the impact of other factors, such as pH, on the sensor signal. NADH and NADPH are autofluorescent, and their concentrations can be measured using fluorescence lifetime imaging microscopy (FLIM) \cite{Yang2021elife,blacker2014separating}. One challenge is to separate NADH signal from NADPH signal, which share the same fluorescence spectrum.

Chemical assays can be used to measure concentrations of biomolecules such as polysaccharides and protein which are critical to cellular composition estimates. Liquid chromatography - mass spectrometry can be used to measure concentrations of a large number of metabolites \cite{park2016metabolite}. However, these methods require destructive sampling and often provide limited spatiotemporal resolution. Nuclear magnetic resonance has been used to measure concentrations of specific metabolites as well \cite{mason1992nmr}. Lastly, stimulated Raman spectroscopy is capable of measuring intracellular metabolite concentrations in living cells with subcellular resolution \cite{oh2019situ}, but it remains a challenge to definitively associate spectral features to specific metabolites.

\subsection{Measuring metabolic fluxes with calorimetry}

As discussed in Sec. \ref{sec:calorimetry}, the heat generated by biological systems is intimately related to their metabolic fluxes. This can be probed by isothermal calorimetry, which measures the heat output of a sample relative to a reference cell at a constant temperature. Isothermal calorimetry is a bulk measurement technique, and does not provide spatially resolved information. Common variants of this technique are listed in Table \ref{tab:calorimetry}.

The temporal resolution of isothermal calorimeters is highly dependent on sample volume. High-sensitivity and low-volume ($\sim$1 $\mathrm{\mu}$L) calorimeters have time resolution on the order of seconds \cite{bae_micromachined_2021,hong_sub-nanowatt_2020,hur_sub-nanowatt_2020}, but the time resolution for larger-volume microcalorimeters is on the order of hours.
Recent devices have achieved sensitivities on the order of 200 pW \cite{bae_micromachined_2021,hong_sub-nanowatt_2020,hur_sub-nanowatt_2020}, which is sufficient to measure the output from single \textit{C. elegans} or \textit{Drosophila} embryos, though single-cell measurements for microbes remain out of reach.

The technical challenges associated with isothermal calorimetry depend strongly on the sample type and amount. In open-chip calorimetry, evaporation can be a significant source of error. For the lowest-volume and most sensitive chip-based calorimeters, with sample volumes less than 1 $\mathrm{\mu}$L, the temperature gradient across the thermopile used to read out a signal is on the order of 1 mK \cite{bae_micromachined_2021}. Thus, even small external temperature gradients, heat sources, or heat losses can significantly impact the measurement.

\begin{table}[h]
\tabcolsep7.5pt
\caption{Selected methods to measure enthalpy flux}
\label{tab:calorimetry}
\begin{center}

\begin{tabular}{|m{0.20\textwidth}|m{0.18\textwidth}|m{0.07\textwidth}|m{0.15\textwidth}|m{0.27\textwidth}|}
\hline
Method & Timescale of measurement & Spatial resolution & Time series possible? & Examples \\ \hline
Isothermal microcalorimetry & varies & none & yes & \textit{E. coli} \cite{braissant_use_2010}, microbial biofilms \cite{astasov-frauenhoffer_isothermal_2012}, \textit{Cutibacterium acnes} \cite{corvec_characterization_2020} \\ \hline
Open-chip calorimetry & varies & none & yes & molecular motors \cite{bae_micromachined_2021}, enzymes \cite{recht_enthalpy_2008}, mouse adipocytes \cite{johannessen_micromachined_2002} \\ \hline
Closed-chip calorimetry & varies & none & yes & bacteria \cite{morais_chip_2014}, \textit{C. elegans} \cite{hur_sub-nanowatt_2020}, \textit{Tetrahymena} \cite{hong_sub-nanowatt_2020} \\ \hline
\end{tabular}

\end{center}
\end{table}

\section{Prior theoretical and experimental approaches to measuring bioenergetic costs}

The flux balance framework and the experimental techniques presented in previous sections are tools that enable measurements of energetic costs of cellular processes. That approach centers on measuring the changes in the ATP fluxes in Eq. \ref{eq:ATPcost_inhibitionexpt} after modulating the activity of a process of interest. In this section, we review previous work that has sought to determine these costs in various systems using a range of other approaches. We discuss these studies' respective challenges and their relation to the flux balance formalism.

\subsection{Estimates and inhibition experiments}

While cellular energy budgets have generally not been probed systematically through experiments, the energetic costs of several key cellular processes have been estimated in different systems. 

A number of attempts have been made to estimate the energetic costs of biosynthesis in growing microbes and developing embryos by leveraging knowledge of the underlying cellular biochemistry. In these calculations, the energetic costs of individual chemical reactions involved in protein, lipid and nucleic acid synthesis, multiplied by the abundances of these different macromolecules in the cell, were summed to arrive at an estimate of the total biosynthesis cost. In the case of protein synthesis, for example, the ATP equivalents required for amino acid activation (1/amino acid) and peptide bond formation (2/amino acid) were accounted for. Such calculations have led some researchers to conclude that  protein synthesis is the major energy consumer in growing microbes, accounting for $\sim$55-65\% of their total energy budget \cite{forrest1971generation,stouthamer1973theoretical, lynch2015bioenergetic}. In contrast, biosynthetic processes have been estimated to account for less than 10\% of the total ATP expenditure in developing \textit{Drosophila} embryos \cite{song_energy_2019}.

Attempts to measure the energetic costs of cellular processes using inhibition experiments have typically focused on characterizing changes in respiratory and/or glycolytic fluxes, without explicitly considering the other ATP fluxes in Eq. \ref{eq:ATPcost_inhibitionexpt}. The results of such experiments suggest that protein synthesis and actin dynamics are the major energy consumers in myoblasts, each accounting for $\sim$20\% of the total ATP production rate \cite{mookerjee2017quantifying}. Similarly, in rabbit brain slices, inhibiting sodium-potassium pumps resulted in a 50\% decrease of oxygen consumption rate, arguing that cation transport is a major energy drain \cite{whittam1962dependence}.

In general, work directly comparing estimates to independent experimental measurements of energetic costs has been limited. However, cases in which these comparisons have been made demonstrate the synergy between these methods. Recent calorimetric measurements revealed oscillations in heat output during zebrafish embryo development, which ceased when the cell cycle was blocked. The authors of that study compared the amplitude and period of the heat oscillation with theoretical estimates of ATP consumption, and argued that they were due to phosphorylation and dephosphorylation reactions which accounted for 1-2\% of the total energy expenditure of the embryo \cite{rodenfels_heat_2019}. In other systems, significant differences have been observed between theoretical estimates and experimental measurements: in neurons, ATP accounting predicted that actin turnover accounts for less than 1\% of energy expenditure \cite{engl2015non}, while inhibition experiments suggested that this process constitutes $\sim$50\% of energy expenditure \cite{bernstein2003actin}. Discrepancies between estimates and measurements highlight the importance of careful examination of assumptions in theoretical calculations and of careful interpretation of process activity modulation experiments. 

Estimates can also provide complementary insights. Chen \textit{et al.} \cite{chen_atp_2015} compared the measured energy cost per beat of a flagellum to two independent theoretical estimates, and found both to be of the same order of magnitude, helping to rationalize the original measurement. In another case, measurements of information processing in the blowfly retina were compared to the minimum energy cost defined by information-theoretic limits. It was found that the measured cost was 5 to 8 orders of magnitude greater than this theoretical minimum \cite{laughlin_metabolic_1998}. This study demonstrated how estimates paired with measurements can provide insight into the origins of energetic costs and the relevance (or lack thereof) of fundamental physical limits. 

Inferring the energetic cost of a specific process from changes in global fluxes after modulating the activity of a process requires knowing the degree and specificity of the change in its activity as well as the potential coupling of different cellular processes. One way to evaluate the accuracy of such a measurement is to compare the inferred cost to measurements of the change of the rate of the process. For example, by measuring incorporation of labeled amino acids, it was shown that the protein synthesis inhibitor cycloheximide completely blocked translation in fish hepatocytes and human hepatoma cells at a concentration of $\sim$25 $\mathrm{\mu}$M, causing a modest decrease in oxygen consumption. However, oxygen consumption continued decreasing as the concentration of cycloheximide was increased, indicating that there were off-target effects at higher concentrations \cite{wieser2001hierarchies}. Experimental measurements may also be cross-validated against estimates based on detailed models of the underlying processes and metabolic pathways \cite{laughlin_metabolic_1998,mahmoudabadi_energetic_2017,lynch_bioenergetic_2015,stouthamer1973theoretical}.

Apart from off-target effects, the intrinsic coupling between different cellular processes could also pose challenges for interpretation of changes in global fluxes. Wieser \textit{et al.} \cite{wieser2001hierarchies} demonstrated that hierarchies of ATP-consuming processes exist in cells: inhibiting mitochondrial respiration and measuring protein synthesis rate and \ce{Na+}/\ce{K+}-ATPase activity in fish hepatocytes and human hepatoma cells revealed that these two processes display different sensitivities to energy limitation in each species. Similar experiments in thymocytes revealed a hierarchy of different processes' sensitivities to inhibition of mitochondrial respiration: protein synthesis was most sensitive, followed by RNA/DNA synthesis and substrate oxidation, \ce{Na+} cycling and \ce{Ca^{2+}} cycling, and finally other ATP consumers and mitochondrial proton leak \cite{buttgereit1995hierarchy}. Application of metabolic control analysis to this data suggested that each ATP consumer exerted negligible influence over the rates of other ATP consumers. However, direct measurements of how modulating the activity of a specific consumer affects the rates of other ATP-consuming process will be required to test these predictions. Given the species and cell type-dependence of the hierarchies of ATP-consuming processes, different systems of interest may need to be individually characterized.

One of the challenges in linking global fluxes with specific fluxes is the potential coupling of the fluxes through different cellular processes. For example, a striking phenomenon of flux homeostasis has been discovered in mouse oocytes, where inhibiting a wide range of energy consuming processes has no effect on the global oxygen consumption rate, despite significantly impacting the metabolic state of the cell \cite{Yang2021elife}. This flux homeostasis implies the coupling of process-specific fluxes inside the cell and an unknown mechanism of flux partitioning that keeps the total flux constant.

\subsection{Using nonequilibrium physics to estimate energetic costs}

An alternative to process activity modulation experiments is to use tools from nonequilibrium physics to estimate energetic costs, or derive limits on costs, using experimentally observable quantities. All irreversible processes consume energy, and nonequilibrium thermodynamics and information theory have been used to place bounds on the associated dissipation \cite{li2019quantifying,battle2016broken,tan2021scale}. Such approaches have also led to proposals for energy-speed-accuracy tradeoffs in diverse biological systems, from single molecular motors to chemotactic bacteria to developing embryos \cite{song_energy_2019,yong2021,lan2021,horowitz2020}.
While a comprehensive treatment of the active field of nonequilibrium thermodynamics is beyond the scope of this review and has been provided elsewhere \cite{fang_nonequilibrium_2019}, we  highlight the Harada-Sasa equality, which connects violation of the fluctuation-response relation to the extent of energy dissipation \cite{harada_equality_2005}. Notably, the calculations required for determining energy dissipation depend solely on experimentally accessible quantities, such as particle trajectories observed through microscopy and cellular rheology measurements. Since only a limited number of degrees of freedom can be experimentally measured, application of the Harada-Sasa equality provides a lower bound on the energetic cost of the underlying process. The predictions of the Harada-Sasa relation have been validated in the case of a particle in an optical trap \cite{toyabe_experimental_2007}.
This equality has also been applied to a number of biologically relevant systems, including molecular motors \cite{harada_fluctuations_2007} and mechanical dissipation inside living cells \cite{fodor_nonequilibrium_2016,bohec_distribution_2019}. For example, this approach was used to determine that the swimming of \textit{Chlamydomonas reinhardtii} had a dissipation rate of $\sim$0.1 fW \cite{jones_stochastic_2021}. Interestingly, this is significantly less than the $\sim$60 fW measured by Chen \textit{et al.} \cite{chen_atp_2015} through monitoring ATP consumption of \textit{C. reinhardtii} flagella \textit{in vitro}. Generalizations of the Harada-Sasa equality have been applied to study free energy dissipation and entropy production in a wider variety of systems, including active matter and nonequilibrium biochemical networks \cite{nardini_entropy_2017,muy_non-invasive_2013,dechant_improving_2021,gingrich_inferring_2017,barato_thermodynamic_2015,wang_landscape_2015}. 

The application of general nonequilibrium relationships, such as the Harada-Sasa equality, is a powerful and elegant approach for studying energy dissipation, but does not leverage prior knowledge about the system of interest. An alternative approach is to develop and test mechanistic,  physics-based models for specific systems. For example, a circuit model of the retinal neuron enabled conversion of membrane potential and conductance measurements into energy consumption rates \cite{laughlin_metabolic_1998}. In another study, Stokes' law was used to estimate the energetic cost of swimming in single-celled organisms, which when compared to the operating cost of flagella revealed that the efficiency of conversion of chemical energy to swimming power was $\sim$0.7\% \cite{schavemaker2022flagellar}.

\section{Open questions}

Despite detailed knowledge of the structure of metabolic networks and the biochemistry of enzymes, relatively little is known about cellular energy expenditures. What are the energetic costs of different cellular processes? How are energy expenditures regulated? Are energy fluxes buffered, and if so, how? How and why do perturbations of energy metabolism lead to different cell biological defects under different circumstances? How do energetic constraints impact cell physiology and the evolution of cellular features? Answering these questions will inform our understanding of bioenergetics, cell biology, and the nonequilibrium physics of living systems. 

Coarse-grained modeling \cite{Yang2021elife,basan2015,trickovic2022, lan2021,niebel2019} may help provide a quantitative description of the mechanisms coupling flux through energy-producing and -consuming pathways. Ultimately, it would be desirable to link such coarse-grained descriptions to the underlying behaviors of the constituent molecular machines, such as respiratory complexes and ATP synthase \cite{abhishek2019,martin2018} to bridge structure, function and physiology across scales.

The flux balance framework presented in this review, culminating in Eq. \ref{eq:ATPcost_inhibitionexpt}, provides a quantitative method to measure energetic costs of cellular processes by studying changes in ATP fluxes upon modulation of process activity. The dynamics of global energetic fluxes such as oxygen consumption rate and heat flux have been measured in many biological contexts, including microbial adaptation, organ function, and embryo development \cite{brettel_microcalorimetric_1981, kooragayala_quantification_2015, neville_novel_2018, koopman_screening-based_2016, astrup_oxygen_1981, engl_non-signalling_2017, robador_nanocalorimetry_2019, peitzsch_real_2008, tourmente_differences_2015,dietz_daily_1998, nagano_temperature-independent_2014, rodenfels_heat_2019,rodenfels_contribution_2020,song_energy_2019, ghosh2022developmental}. However, the contributions of specific energy-consuming processes such as biosynthesis, ion pumping, and cytoskeletal remodeling to changes in global fluxes remain unclear. Systematic study of bioenergetic fluxes and their coupling will be required to measure the costs of cell biological processes.

%Disclosure
\section*{DISCLOSURE STATEMENT}
The authors are not aware of any affiliations, memberships, funding, or financial holdings that might be perceived as affecting the objectivity of this review. 

% Acknowledgements
\section*{ACKNOWLEDGMENTS}
We thank Yong Hyun Song, Peter Foster, and Piyush Nanda for discussions and comments. This work was supported by National Science Foundation awards PHY-2013874 and MCB-2052305 to D.J.N.

\printbibliography

\end{document}